\documentclass[referee]{raa}            

\usepackage{graphicx,times}             
\usepackage{natbib}
\usepackage{amssymb,amsmath}
\usepackage{hyperref}
\bibpunct{(}{)}{;}{a}{}{,}

\newcommand{\RNum}[1]{\uppercase\expandafter{\romannumeral #1\relax}}
\hypersetup{colorlinks = true, linkcolor = red, anchorcolor = red, citecolor = blue, filecolor = red, urlcolor = red, backref = page}

\begin{document}

\title{Constraints on axionlike particles with different magnetic field models from the PKS 2155-304 energy spectrum}

   \volnopage{Vol.0 (20xx) No.0, 000--000}      
   \setcounter{page}{1}          

  \author{Jia Bu
      \inst{1}
   \and Ya-Ping Li
      \inst{2}
      }\
      \institute{Department of Astronomy, Xiamen University, Xiamen 361005, China; {\it 19820141152948@stu.xmu.edu.cn}\\
        \and
                 Shanghai Astronomical Observatory, Shanghai 200030, China\\
\vs\no
   {\small Received~~2018~~month day; accepted~~year~~month day}}

\abstract{Axion Like Particles (ALPs) are one promising kind of dark matter candidate particles that are predicted to couple with photons in the presence of magnetic fields. The oscillations between photons and ALPs travelling in the magnetic fields have been used to constrain ALP properties. In this work, we obtain some new constraints on the ALP mass $m_{\rm a}$ and the photon-ALP coupling constant $g$ with two different magnetic field models through TeV photons from PKS 2155-304. One is the discrete-$\varphi$ model that the magnetic field has the orientation angle $\varphi$ changes discretely and randomly from one coherent domain to the next, another is the linearly-continuous-$\varphi$ model that the magnetic field orientation angle $\varphi$ varies continuously across neighboring coherent domains. For the discrete-$\varphi$ model, we can obtain the best constraints on the ALP mass $m_{1}=m_{\rm a}/(1\  {\rm neV})=0.1$ and on the photon-ALP coupling constant $g_{11}=g/(10^{-11}\ {\rm GeV^{-1}})=5$, the reasonable range of the ALP mass $m_{1}$ is 0.08 $\thicksim$ 0.2 when $g_{11}$=5, and the only reasonable value of the photon-ALP coupling constant is $g_{11}$=5 when $m_{1}$=0.1. For the linearly-continuous-$\varphi$ model, we can obtain the best constraints on the ALP mass $m_{1}=0.1$ and on the photon-ALP coupling constant $g_{11}=0.7$, the reasonable range of the ALP mass $m_{1}$ is 0.05 $\thicksim$ 0.4 when $g_{11}$=0.7, and the reasonable range of the photon-ALP coupling constant $g_{11}$ is 0.5 $\thicksim$ 1 when $m_{1}$=0.1. All the results are consistent with the upper bound ($g<6.6\times10^{-11}\ {\rm GeV^{-1}}$, i.e. $g_{11}<6.6$) set by the CAST experiment.
}

   \authorrunning{Jia Bu, Ya-Ping Li }            
   \titlerunning{Constraints on axionlike particles with different magnetic field models}  

   \maketitle

\section{INTRODUCTION}
Axions are predicted by the Peccei-Quinn mechanism, which is a noticeable explanation to solve the strong CP problem in QCD (\citealt{Peccei1977}). In a more generic way, Axion-Like Particles (ALPs) appear in extensions of the standard model of particle physics (\citealt{Jaeckel2010}). In the presence of an external magnetic field $\mathbf{B}$, ALPs (represented by the field $a$) have a general property that they can couple with photons (represented by $\mathbf{E}$) through the interaction Lagrangian $\mathcal{L}=ga\mathbf{E}\mathbf{\cdot}\mathbf{B}$. A photon can oscillate into an ALP and vise versa. Such photon-ALP oscillations have been used to explain lots of astrophysical phenomena, or to constrain the properties of ALPs (\citealt{Mirizzi2008}). For example, the apparent dimming of supernovae (\citealt{Ostman2005, Mirizzi2005, Csaki2002}), the spectral distortions of the cosmic microwave background (\citealt{Csaki2015, Dias2014}), and the anomalous lack of opacity to gamma rays of the Universe: HESS, MAGIC and Fermi have detected high energy gamma ray photons in the TeV range from distance Active Galactic Nuclei (AGNs) (\citealt{H.E.S.S.2013, Mazin2007, Aharonian2006, Ackermann2012,  MAGIC2008}). Before reaching the earth, these photons will suffer significant attenuation because electron-positron pair production on the extragalactic background infrared radiation. There is a possible interpretation for this transparency phenomenon, i.e. due to the photon-ALP mixing, the radiation from AGNs travel in the form of ALPs without producing pairs in most of the distance, and convert back to the photons when they arriving the earth (\citealt{Mirizzi2009, Burrage2009, Simet2008}).

The magnetic field structure may strongly affect photon-ALP propagation in the universe. The most previous studies adopted a magnetic field model that the path of propagation is divided into many coherent domains, each has a uniform magnetic field and the same size $l$. In this model, the magnetic field has the orientation angle (represented by $\varphi$) changes discretely and randomly from one domain to the next. Based on this model, \citet{Grossman2002} derived a formula for the photon-to-ALP conversion probability through plenty of domains, and this formula has been widely used in lots of previous work. But recently \citet{Wang2016} adopted another magnetic field model that the magnetic field orientation angle $\varphi$ varies continuously across neighboring domains. \citet{Wang2016} showed that qualitative significantly different result for the photon-to-ALP conversion probability compared with which obtained in the discrete-$\varphi$ model.

In this work, we compare the two different results of the photon-ALP propagation for two different magnetic field models mentioned above, and obtain some new constraints on ALP properties based on photon-to-ALP conversion probability through TeV photons from a distant AGN PKS 2155-304.

This paper is organized as follows. In section \RNum{2}, we describe the fitting of the broadband spectral energy distribution (SED) of PKS 2155-304 with a one-zone synchrotron self-Compton (SSC) model. In section \RNum{3}, we briefly summarize the two different results of the
photon-ALP propagation for the two different magnetic field models. In section \RNum{4}, we compare the constrains of ALP properties for the two different magnetic field models through fitting the survival probabilities of the Tev photons of PKS 2155-304. Conclusions and discussion are presented in the last section.

\section{ OBSERVATION AND FITTING OF PKS 2155-304}

   \begin{figure}
   \centering
   \includegraphics[width=0.7\textwidth, angle=0]{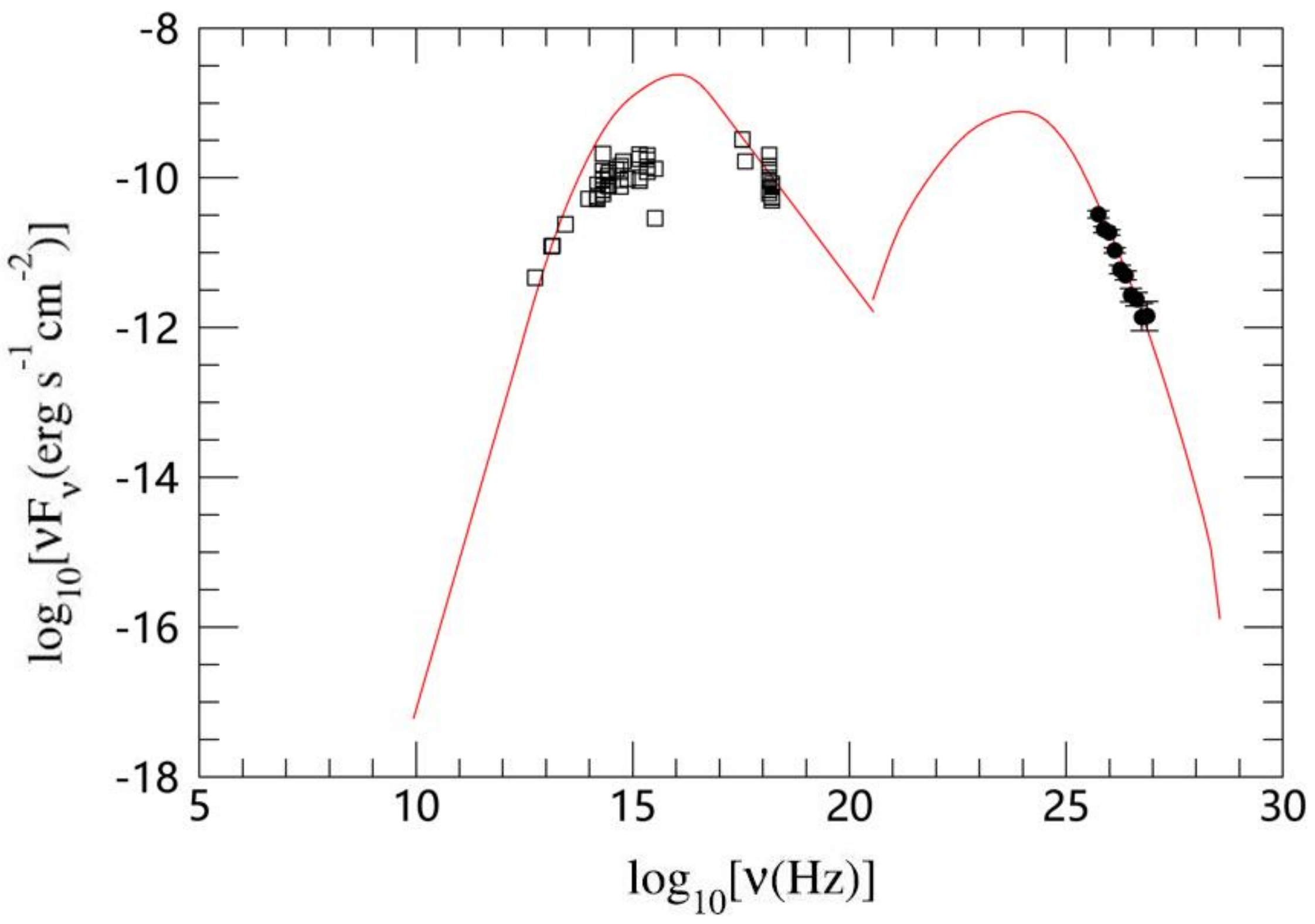}
   \caption{The fitting of the broadband SED of PKS 2155-304, where the red solid line shows the one-zone SSC fitting of the SED of PKS 2155-304, the black open squares show the broadband observed data from NED, the black filled circles show the Tev data from the HESS observations (\citealt{Aharonian2005b}) }
   \label{fig:fitting}
   \end{figure}
Recent studies reveal that an irregular local energy spectrum of PKS 2155-304 has been used to constrain the photon-ALP coupling (\citealt{Abramowski2013}). PKS 2155-304 is a powerful and well-studied TeV gamma-ray source (\citealt{Abramowski2010, Aharonian2009a, Aharonian2007, Aharonian2005a}). This BL Lac may offer the possibility for the conversion of photon-ALP in the magnetic field which along the path of propagation (\citealt{Falomo1993, Smith1995, Abramowski2013}). So PKS 2155-304 is a good target for the ALP research at high energies (\citealt{Horns2012}).

We have collected the broadband observed data from the NASA/IPAC Extragalactic Database (NED), along with the Tev data from the HESS observations. We fit the broadband SED of PKS 2155-304 with a one-zone SSC model developed by \cite{Chen2017}. In this one-zone SSC model, there are nine independent parameters to depict the broadband SED. Three parameters characterize the global properties of the emitting blob: $\delta$ - the Doppler factor, $B$ - the uniform magnetic field, and $R$ - the radius of the emitting blob. The other six parameters characterize the distribution and the physical properties of the high-energy particles: $p_{1}$, $p_{2}$ - the spectral indexes of the electron energy distribution at lower and higher energies respectively, $\gamma_{\rm max}$, $\gamma_{\rm min}$, and$\gamma_{0}$ - the maximum, minimum, and break Lorentz factors of the electron energy distribution, $N_{0}$ - the density factor. In this model, the synchrotron + SSC emissions can produce the whole SED. We can adjust the nine free independent parameters as mentioned above to fit the SED of PKS 2155-304. Fig.~\ref{fig:fitting} presents our fitting, and the corresponding parameters are listed in Table~1. In Fig.~\ref{fig:fitting}, the red solid line shows the one-zone SSC fitting of the SED of PKS 2155-304, the black open squares show the broadband observation data from NED, and the black filled circles show the Tev data from the HESS observations (\citealt{Aharonian2005b}).

\begin{table}
\begin{center}
\caption[]{The One-Zone SSC Model Parameters for PKS 2155-304}\label{tab:fittting}
\begin{tabular}{ccccccccc}
\hline
\hline
$\delta$ &  B    &      R         &  $N_{0}$   & $\gamma_{\rm min}$ & $\gamma_{0}$ & $\gamma_{\rm max}$ & $p_{1}$ & $p_{2}$\\
         &  (G)  &     (cm)       & (${\rm cm}^{-3}$)&                &              &                &         &        \\ \hline
   35    & 1.2 & $8\times10^{15}$ &   6000     &      1100      &     9000     &   $10^{7}$     &   2     & 4.54   \\  \hline \hline
\end{tabular}
\end{center}
\tablecomments{\textwidth}{$\delta$ is the Doppler factor, $B$ is the uniform magnetic field, $R$ is the radius of the emitting blob,   $N_{0}$ is the density factor, $\gamma_{\rm min}$,$\gamma_{0}$ and $\gamma_{\rm max}$ are the minimum, break and maximum Lorentz factors of the electron energy distribution, $p_{1}$ and $p_{2}$ are the spectral indexes of the electron energy distribution at lower and higher energies respectively.}
\end{table}

\section{PHOTON-ALP PROPAGATION FOR TWO MAGNETIC FIELD MODELS}
We briefly summarize the two different results of the photon-ALP propagation for two different models of magnetic field which along the path of propagation below, and refer readers to \citet{Kuo1989, Grossman2002, Wang2016} for details.

In a Cartesian coordinates X-Y-Z (the Z-axis is along the direction of propagation), if the angular frequency $\omega$ or energy $\varepsilon$ is given (for $\mathbf{E}$,$a$ $\propto$ $e^{i\omega t}$ ), the evolution equation of ALP field $a$ and the photon electric field $\mathbf{E}$ can take the form
\begin{equation}
i\left( \begin{array}{c}
a^{\prime} \\
E_{x}^{\prime} \\
E_{y}^{\prime}
\end{array} \right) = \left( \begin{array}{ccc}
\omega+\Delta_{\rm a} & \Delta_{\rm M}\cos\varphi & \Delta_{\rm M}\sin\varphi \\
\Delta_{\rm M}\cos\varphi & \omega+\Delta_{\rm pl} & 0 \\
\Delta_{\rm M}\sin\varphi & 0 & \omega+\Delta_{\rm pl}
\end{array} \right)
\left( \begin{array}{c}
a \\
E_{x} \\
E_{y}
\end{array} \right),
\end{equation}
where the superscript $\prime$ represents $d/dz$, and $\varphi$ is the orientation angle of the magnetic field $\mathbf{B}$, i.e.$\varphi$ is the angle between $\mathbf{B}_{\rm tr}$ (the projection of $\mathbf{B}$ in the XY-plane) and the X-axis. If we define dimensionless quantities (in units of $c=\hbar=1$),
\begin{equation}
\begin{array}{l@{\quad\quad}l}
m_{1}=m_{\rm a}/(1\ {\rm neV}), \\
\varepsilon_{1}=\varepsilon/(1\ {\rm TeV}), \\
g_{11}=g/(10^{-11}\  {\rm GeV}^{-1}), \\
B_{1}=B_{\rm tr}/(1\ {\rm nG}).
\end{array}
\end{equation}

The parameter $\Delta_{\rm a}$ is related to the ALP mass and the parameter $\Delta_{\rm M}$ is related to the photon-ALP coupling constant, they are given by
\begin{equation}
\Delta_{\rm a}=-\frac{m_{\rm a}^{2}}{2\omega}=-7.83\times10^{-2}\varepsilon_{1}^{-1}m_{1}^{2}\ {\rm Mpc}^{-1}, \\
\end{equation}
and
\begin{equation}
\Delta_{\rm M}=\frac{1}{2}gB_{\rm tr}=4.63\times10^{-3}g_{11}B_{1}\ {\rm Mpc}^{-1},
\end{equation}
where $m_{\rm a}$ is the ALP mass, $\varepsilon$ is the photon energy, $g$ is the photon-ALP coupling constant.
The plasma parameter $\Delta_{\rm pl}=-\omega_{\rm pl}^{2}/(2\omega)=-1.11\times10^{-11}\varepsilon_{1}^{-1}(n_{\rm e}/10^{-7}\ {\rm cm}^{-3})\ {\rm Mpc}^{-1}$, where $\omega_{\rm pl}$ is the electron plasma frequency and $n_{\rm e}$ is the electron density.

The photon-ALP evolution can be obtained by integrating Eq.(10) along the ray for a given magnetic field structure. We assume that the path of propagation is divided into many coherent domains, each has a uniform magnetic field and the same size $l$.

For the model that the magnetic field has the orientation angle $\varphi$ changing discretely and randomly from one domain to the next, \citet{Grossman2002} derived an analytic formula for the mean value of the photon-to-ALP conversion probability (represented by $P_{\rm G}$) after propagating through $N$ domains(over distance $D=Nl$), and this formula has been widely used in lots of previous work:
\begin{equation}
P_{\rm G}=\frac{1}{3}(1-e^{-3NP_{0}/2}),
\end{equation}
where
\begin{equation}
P_{0}=\frac{\Delta_{\rm M}^{2}}{(\Delta k/2)^{2}}\sin^{2}(\Delta kD/2),
\end{equation}
and $\Delta k=\sqrt{\Delta_{\rm a}^{2}+4\Delta_{\rm M}^{2}}$.

For the model that the magnetic field orientation angle $\varphi$ varies linear-continuously across neighboring domains, \citet{Wang2016} obtained a numerical result shows that this continuous-$\varphi$ model can generate completely different photon-to-ALP conversion probability compared to the discrete-$\varphi$ model. The mean photon-to-ALP conversion probability (represented by $P_{\rm W}$) after propagating through $N$ domains (over distance $D=Nl$) is
\begin{equation}
P_{\rm W} \simeq 0.123\frac{\Delta_{\rm M}^{2}}{\Delta_{\rm a}^{2}}[1-\cos(N\Delta_{\rm a}l)]+\sigma_{\rm A}^{2}N\Delta_{\rm M}^{2}l^{2}.
\end{equation}
It is notable that the validity of Eq.(16) requires $P_{\rm W}\ll1$ (\citealt{Wang2016}).

\section{CONSTRAINS ON ALP PROPERTIES}
Based on the fitting of the broadband SED of PKS 2155-304 (see Fig.~\ref{fig:fitting}), we can calculate the survival probabilities (represented by $P_{\rm S}$) of the Tev photons of PKS 2155-304: $P_{\rm S}=F_{\rm obs}/F_{\rm source}$, where $F_{\rm obs}$ is the observed fluxes of Tev photons, $F_{\rm source}$ is the fluxes of the Tev photons before propagation in the universe, i.e. the fitting values in SED of PKS 2155-304.

We can also get the Tev photons survival probabilities through the photons-to-ALPs conversion probabilities, because the survival probability plus the conversion probability equals 1 for one single photon.

For the model that the magnetic field has the orientation angle $\varphi$ changing discretely and randomly from one domain to the next, the Tev photon survival probability (represented by $P_{\rm S,G}$) is:
\begin{equation}
P_{\rm S,G}=1-P_{\rm G}=\frac{2}{3}+\frac{1}{3}e^{-3NP_{0}/2}.
\end{equation}

For the model that the magnetic field orientation angle $\varphi$ varies linear-continuously across neighboring domains, the Tev photon survival probability (represented by $P_{\rm S,W}$) is:
\begin{eqnarray}
P_{\rm S,W}&=&1-P_{\rm W}\simeq 1-0.123\frac{\Delta_{\rm M}^{2}}{\Delta_{\rm a}^{2}}[1-\cos(N\Delta_{\rm a}l)]-\sigma_{\rm A}^{2}N\Delta_{\rm M}^{2}l^{2}.
\end{eqnarray}

Note that the magnetic field structure around the source and in the intergalactic medium (IGM) are very different. Typically, the intergalactic magnetic field has an upper limit of a few nG and the coherent domain size is on the order of a few Mpc , but the strength of the magnetic field around the source is about 0.1$\thicksim$1G and coherent domain size is about 0.1$\thicksim$ a few pc (\citealt{Conde2009, Grasso2001, Grossman2002, Ade2015}). For PKS 2155-304, in our calculation, we adopt the following scheme: the strength of the magnetic field around the source (represented by $B_{\rm sou}$) we adopt is about $1.2\ {\rm G}$, which is  the value adopted in our SSC model (see Table~1), and the coherent domain size around the source (represented by $l_{\rm sou}$) we adopt is about $1\ {\rm pc}$, which is  the distance from the central black hole to the broad line region (BLR); the strength of the intergalactic magnetic field (represented by $B_{\rm int}$) we adopt is about $1\ {\rm nG}$, and the intergalactic coherent domain size (represented by $l_{\rm int}$) we adopt is about $1\ {\rm Mpc}$. We have listed all the values in Table~2.

\begin{table}
\begin{center}
\caption[]{Parameters used to calculate the total photon-axion conversion both in the magnetic field around the source and in the intergalactic magnetic field.}\label{tab:best-fittting}
\begin{tabular}{c|c|c|c}
\hline
\hline
              &  Parameter& discrete-$\varphi$ model & linearly-continuous-$\varphi$ model   \\ \hline
Source        &  $B_{\rm sou}$  &  1.2 G    &  1.2 G      \\
parameters    &  $l_{\rm sou}$  &  1 pc     &  1 pc       \\  \hline
Intergalactic &  $B_{\rm int}$  &  1 nG     &  1 nG       \\
parameters    &  $l_{\rm int}$  &  1 Mpc    &  1 Mpc      \\  \hline
ALP parameters&  $m_{\rm a}$    &  $0.1\  {\rm neV}$ &  $0.1\  {\rm neV}$   \\
(best fitting)&  $g$          &$5\times10^{-11}\  {\rm GeV}^{-1}$ &$0.7\times10^{-11}\  {\rm GeV}^{-1}$\\ \hline \hline
\end{tabular}
\end{center}
\tablecomments{\textwidth}{$B_{\rm sou}$ represents the strength of the magnetic field around the source, $l_{\rm sou}$ represents the coherent domain size around the source, $B_{\rm int}$ represents the strength of the intergalactic magnetic field, $l_{\rm int}$ represents the intergalactic coherent domain size, $m_{\rm a}$ represents the ALP mass, $g$ represents the photon-ALP coupling constant.}
\end{table}

If the survival probability of the Tev photon which propagates through the magnetic field around the source is represented by $P_{\rm S,sou}$, and the survival probability propagating through the intergalactic magnetic field is represented by $P_{\rm S,int}$, the total survival probability can be written as
\begin{equation}
P_{\rm S}=P_{\rm S,sou} \times P_{\rm S,int}.
\end{equation}

It is notable that the generated ALPs in the magnetic field around the source would be partially converted back to photons in the intergalactic magnetic field. However, based on our calculations, we find that this effect is too small for the total survival probability. So we can neglect it.

For the two different models, we have:
\begin{eqnarray}
P_{\rm S,G}&=&P_{\rm S,G,sou} \times P_{\rm S,G,int}\\
           &=&\left(\frac{2}{3}+\frac{1}{3}e^{-3NP_{0}/2}\right)_{\rm sou}\times\left(\frac{2}{3}+\frac{1}{3}e^{-3NP_{0}/2}\right)_{\rm int},\nonumber
\end{eqnarray}
and
\begin{eqnarray}
P_{\rm S,W}&=&P_{\rm S,W,sou} \times P_{\rm S,W,int}\\
           &=&\left(1-0.123\frac{\Delta_{\rm M}^{2}}{\Delta_{\rm a}^{2}}\left[1-\cos(N\Delta_{\rm a}l)\right]-\sigma_{\rm A}^{2}N\Delta_{\rm M}^{2}l^{2}\right)_{\rm sou}\nonumber\\
           &\times&\left(1-0.123\frac{\Delta_{\rm M}^{2}}{\Delta_{\rm a}^{2}}\left[1-\cos(N\Delta_{\rm a}l)\right]-\sigma_{\rm A}^{2}N\Delta_{\rm M}^{2}l^{2}\right)_{\rm int}.\nonumber
\end{eqnarray}

For the expressions (11) and (12), we can adjust the parameters $m_{1}$ and $g_{11}$ to make a best fitting of the survival probabilities of the Tev photons of PKS 2155-304.
\begin{figure}
   \centering
   \includegraphics[height=6.5cm,width=9cm]{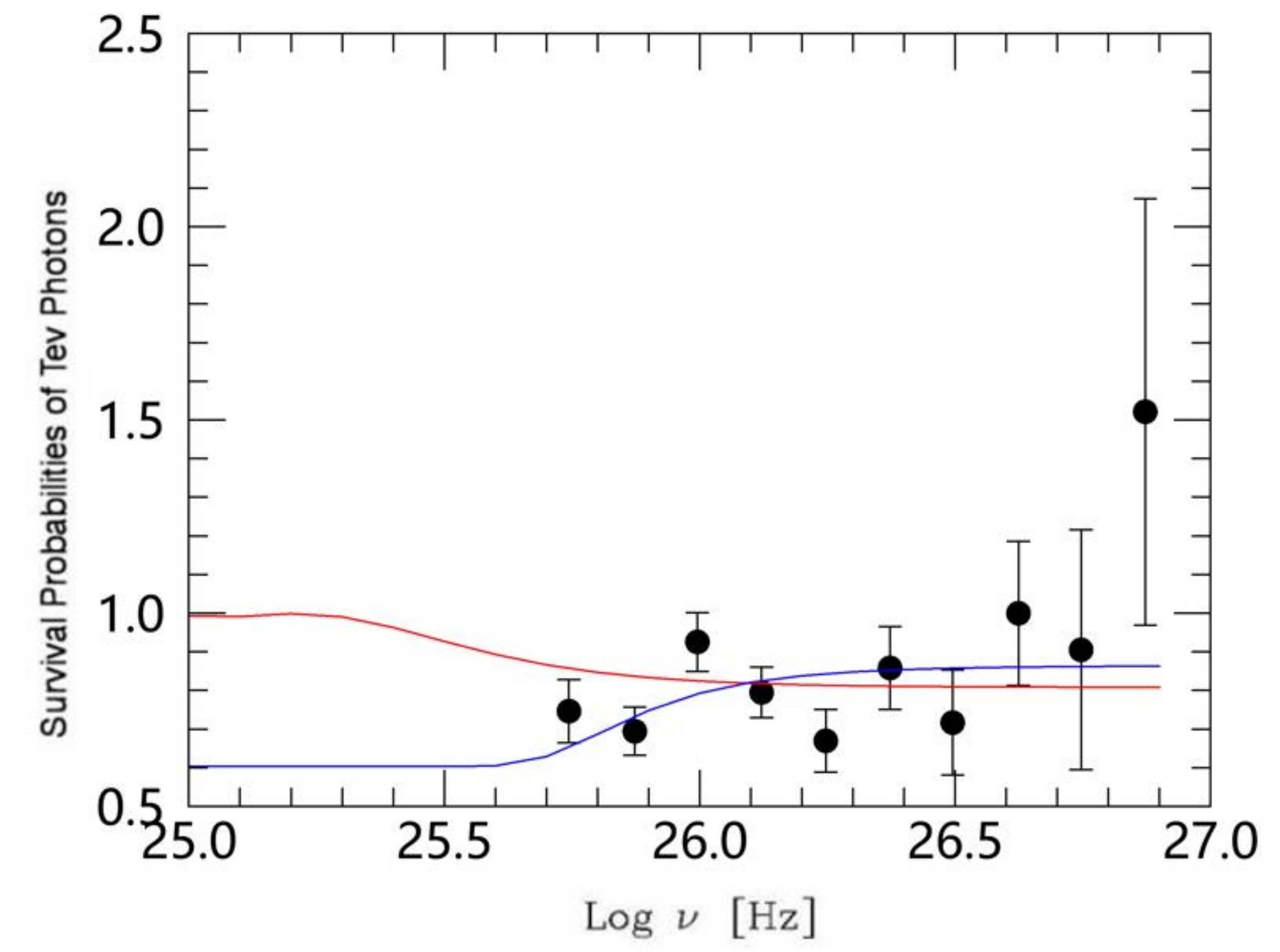}
   \caption{The best fitting of survival probabilities of the Tev photons of PKS 2155-304 for the two different model. Where the black
filled circles show the survival probabilities ($P_{\rm S}$) of all the Tev photons. The blue solid line shows the best fitting for the discrete-$\varphi$ model ($P_{\rm S,G}$), and the corresponding parameter values are: $m_{1}=0.1$, $g_{11}=5$. The red solid line shows the best fitting for the linearly-continuous-$\varphi$ model ($P_{\rm S,W}$), and the corresponding parameter values are:$m_{1}=0.1$, $g_{11}=0.7$. The highest black filled circle is not considered in our fitting, because its error is too big and the survival probability it represents is greater than one, which is unphysical.}
   \label{fig:best-fitting}
   \end{figure}

In Fig.~\ref{fig:best-fitting}, we present the survival probabilities ($P_{\rm S}$) of all the Tev photons, which are showed by the black filled circles. We also present the best fitting curves for the discrete-$\varphi$ model ($P_{\rm S,G}$) and for the linearly-continuous-$\varphi$ model ($P_{\rm S,W}$) respectively. For Eq.(11), when the ALP mass $m_{1}$=0.1 and the photon-ALP coupling constant $g_{11}$=5, we can obtain the best fitting curve of $P_{\rm S,G}$, which is showed by the blue solid line. For Eq.(14), when $m_{1}$=0.1 and $g_{11}$=0.7, we can obtain the best fitting curve of $P_{\rm S,W}$, which is showed by the red solid line. It is notable that we didn't consider the highest black filled circle in Fig.~\ref{fig:best-fitting} in our fitting, because its error is too big and the survival probability it represents is greater than one, which is unphysical.

\begin{figure*}[t]
\center
\includegraphics[height=0.25\textheight,width=0.49\textwidth]{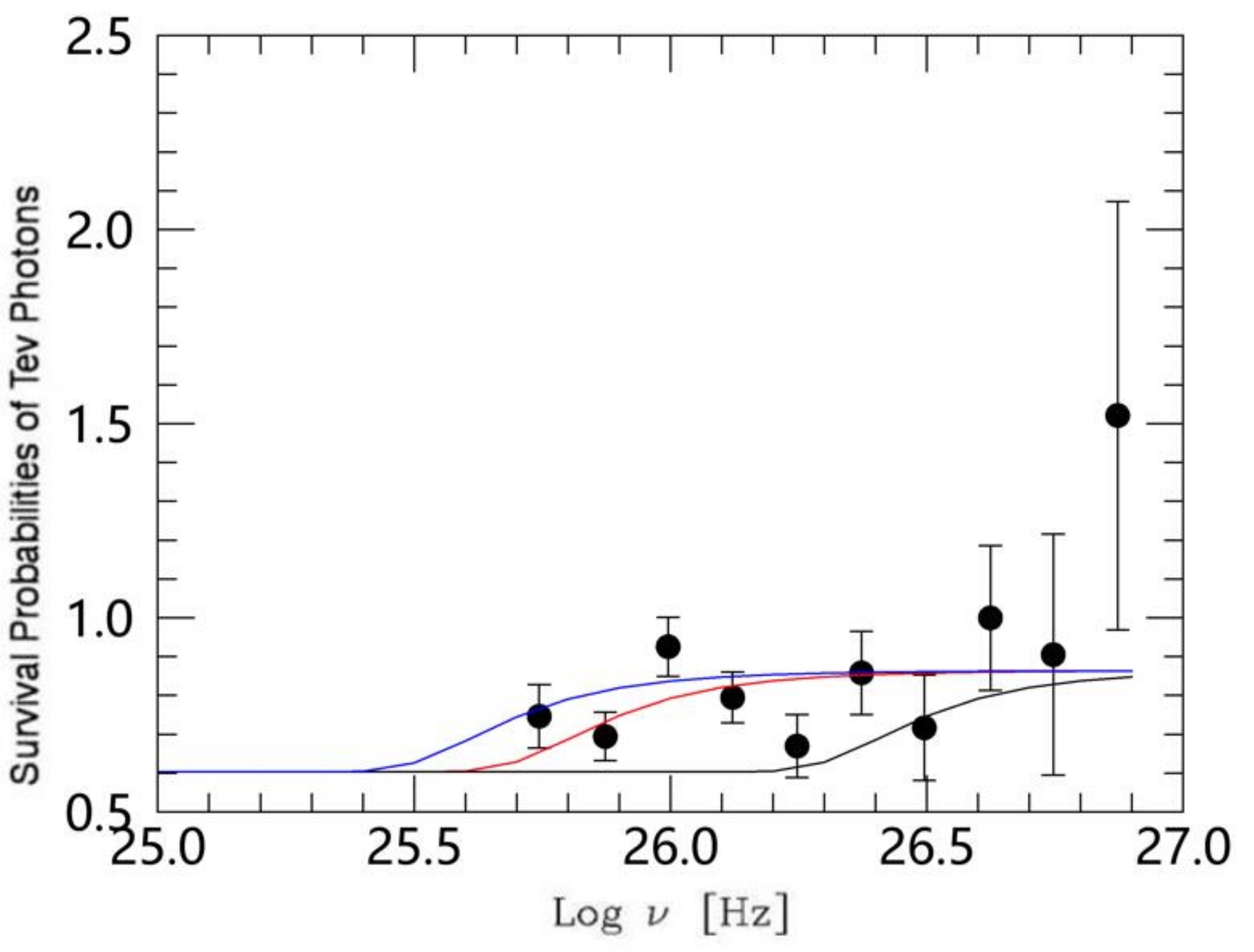}
\includegraphics[height=0.25\textheight,width=0.49\textwidth]{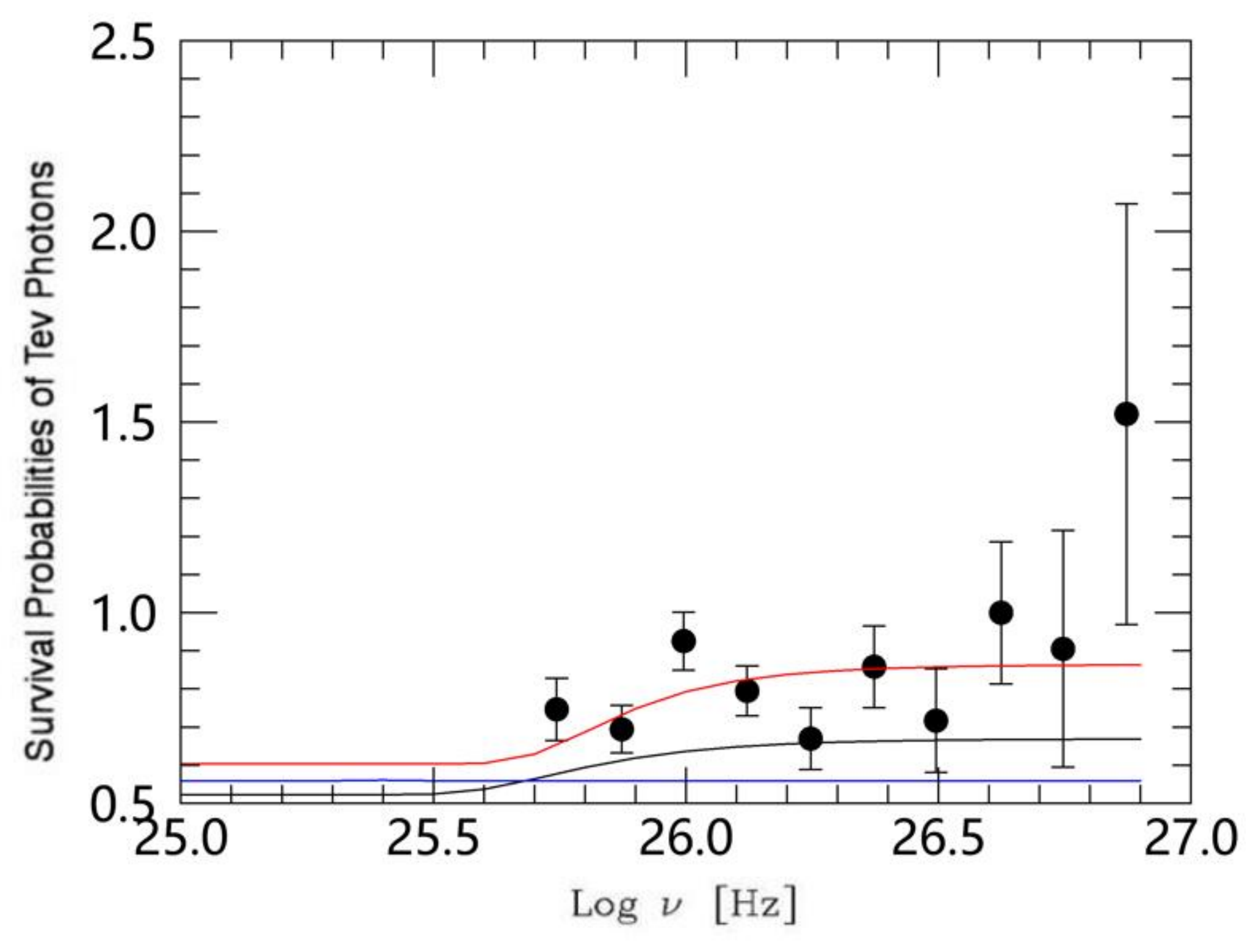}
\caption{A few typical fitting curves for the discrete-$\varphi$ model. The left panel shows the different fitting curves of $P_{\rm S,G}$ with different $m_{1}$ when $g_{11}$=5: the blue solid line shows $m_{1}$=0.08, the red solid line shows $m_{1}$=0.1, and the black solid line shows $m_{1}$=0.2. The right panel shows the different fitting curves of $P_{\rm S,G}$ with different $g_{11}$ when $m_{1}=0.1$: the red solid line shows $g_{11}$=5, the black solid line shows $g_{11}$=10, and the blue solid line shows that $g_{11}$ takes other values.}
\label{fig:Gcompare}
\end{figure*}

\begin{figure*}[t]
\center
\includegraphics[height=0.25\textheight,width=0.49\textwidth]{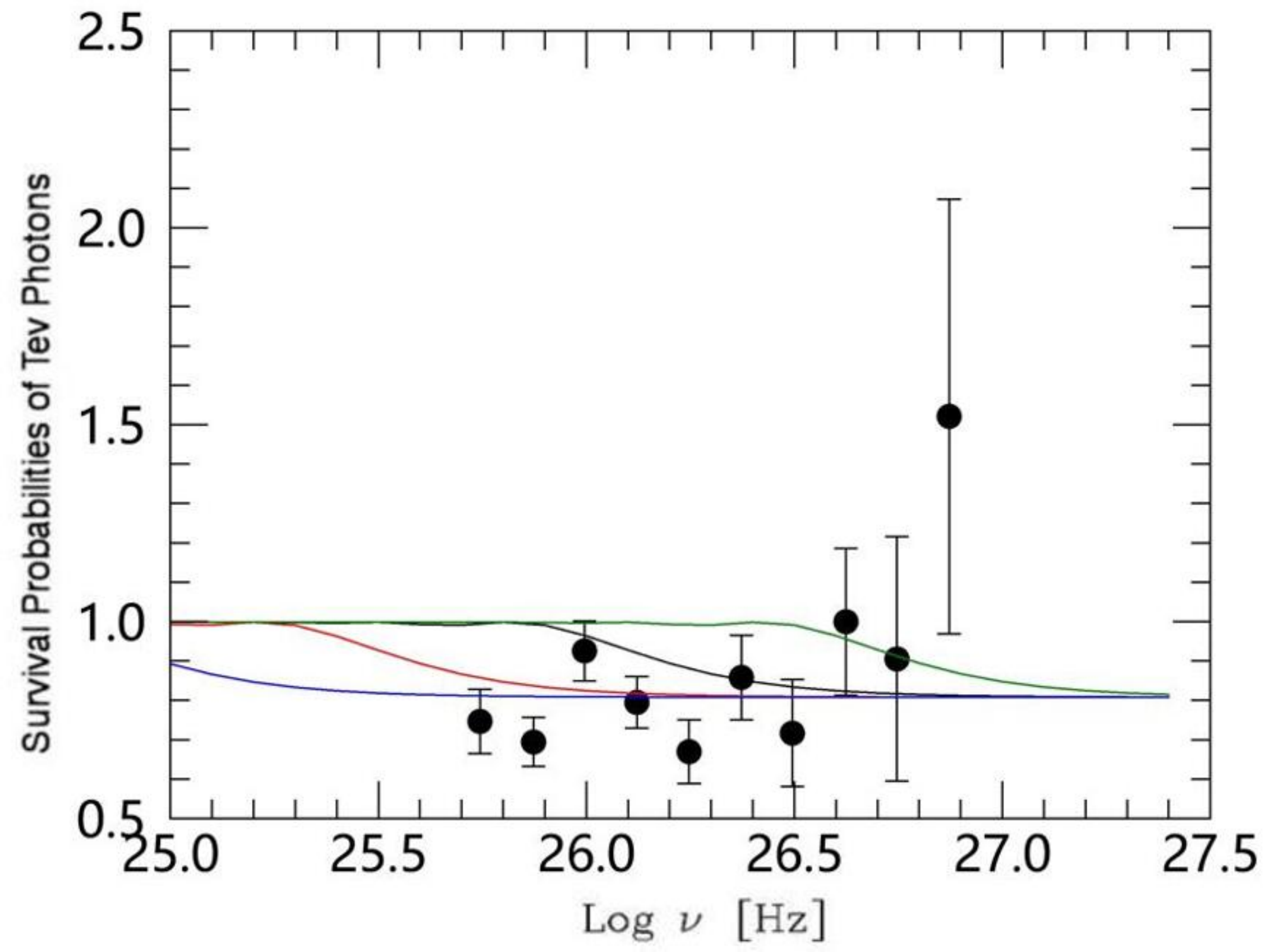}
\includegraphics[height=0.25\textheight,width=0.49\textwidth]{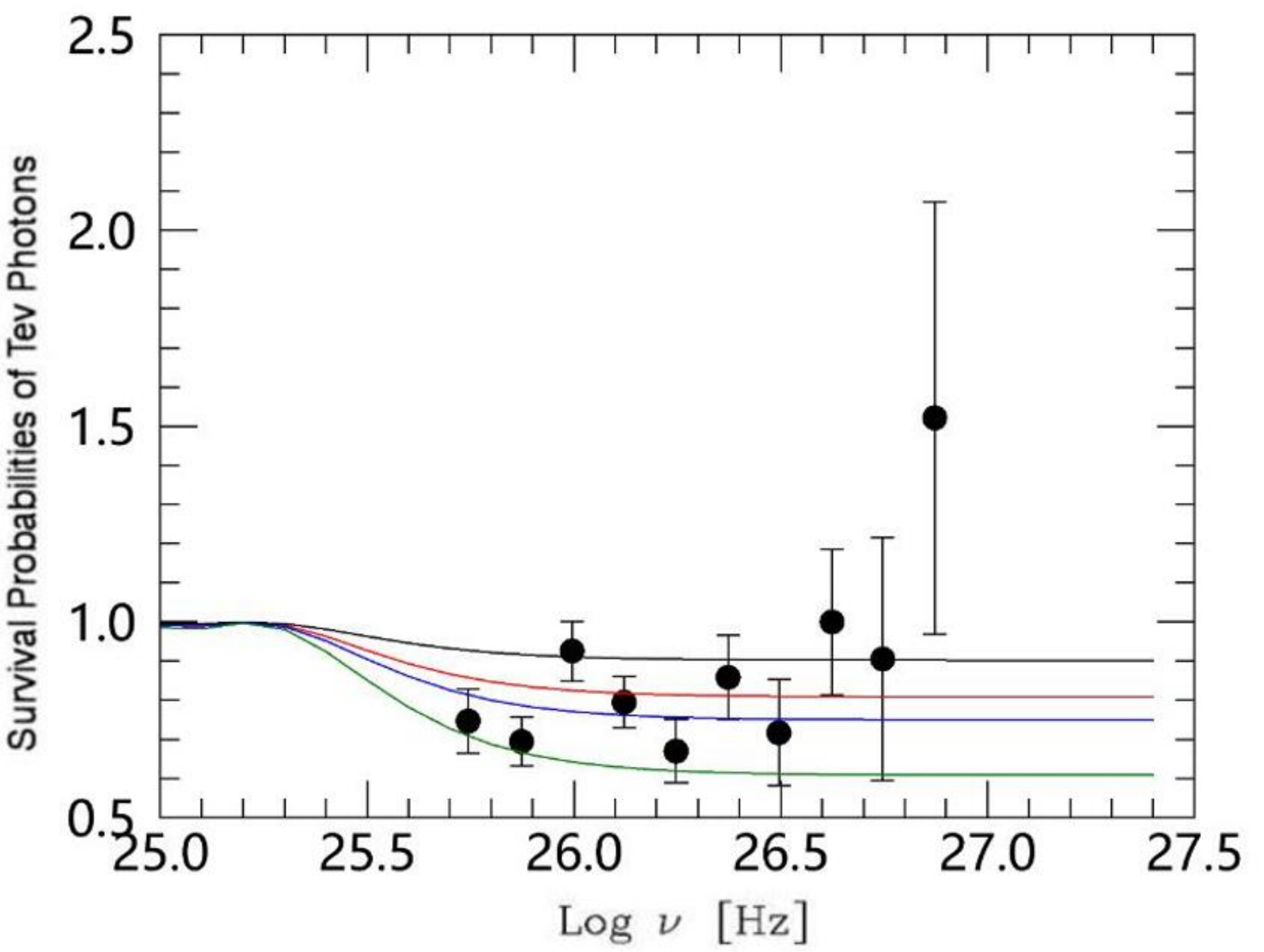}
\caption{A few typical fitting curves for the linearly-continuous-$\varphi$ model. The left panel shows the different fitting curves of $P_{\rm S,W}$ with different $m_{1}$ when $g_{11}$=0.7: the blue solid line shows $m_{1}$=0.05, the red solid line shows $m_{1}$=0.1, the black solid line shows $m_{1}$=0.2, and the green solid line shows $m_{1}$=0.4. The right panel shows the different fitting curves of $P_{\rm S,W}$ with different $g_{11}$ when $m_{1}$=0.1: the black solid line shows $g_{11}$=0.5, the red solid line shows $g_{11}$=0.7, the blue solid line shows $g_{11}$=0.8, and the green solid line shows $g_{11}$=1.}
\label{fig:Wcompare}
\end{figure*}

For the discrete-$\varphi$ model, we present a few typical fitting curves in Fig.~\ref{fig:Gcompare}. The left panel shows the different fitting curves of $P_{\rm S,G}$ with different $m_{1}$ when $g_{11}$=5: the blue solid line shows $m_{1}$=0.08, the red solid line shows $m_{1}$=0.1, and the black solid line shows $m_{1}$=0.2. So the reasonable range of the ALP mass $m_{1}$ is 0.08 $\thicksim$ 0.2 when $g_{11}$=5. The right panel shows the different fitting curves of $P_{\rm S,G}$ with different $g_{11}$ when $m_{1}$=0.1: the red solid line shows $g_{11}$=5, the black solid line shows $g_{11}$=10, and the blue solid line shows that $g_{11}$ takes other values. So the only reasonable value of the photon-ALP coupling constant is $g_{11}$=5 when $m_{1}$=0.1. These results imply that, in the energy range $10^{25} -10^{27}\ {\rm Hz}$, for the discrete-$\varphi$ model, the Tev photon survival probabilities $P_{\rm S,G}$ are very sensitive to the ALP mass $m_{1}$, but are not sensitive to the photon-ALP coupling constant $g_{11}$.

For the linearly-continuous-$\varphi$ model, we also present a few typical fitting curves in Fig.~\ref{fig:Wcompare}. The left panel shows the different fitting curves of $P_{\rm S,W}$ with different $m_{1}$ when $g_{11}$=0.7: the blue solid line shows $m_{1}$=0.05, the red solid line shows $m_{1}$=0.1, the black solid line shows $m_{1}$=0.2, and the green solid line shows $m_{1}$=0.4. So the reasonable range of the ALP mass $m_{1}$ is 0.05 $\thicksim$ 0.4 when $g_{11}$=0.7. The right panel shows the different fitting curves of $P_{\rm S,W}$ with different $g_{11}$ when $m_{1}$=0.1: the black solid line shows $g_{11}$=0.5, the red solid line shows $g_{11}$=0.7, the blue solid line shows $g_{11}$=0.8, and the green solid line shows $g_{11}$=1. So the reasonable range of the photon-ALP coupling constant $g_{11}$ is 0.5 $\thicksim$ 1 when $m_{1}=0.1$. These results imply that, in the energy range $10^{25}Hz-10^{27}Hz$, for the linearly-continuous-$\varphi$ model, the Tev photon survival probabilities $P_{\rm S,W}$ are very sensitive to the ALP mass $m_{1}$, and are also very sensitive to the photon-ALP coupling constant $g_{11}$.

It is difficult to explain why there is only one reasonable value of the photon-ALP coupling constant (i.e. $g_{11}$=5) when $m_{1}$=0.1 for the discrete-$\varphi$ model, but for the two models, the best-fitting parameters of ALPs and the reasonable ranges of the parameters of ALPs which are based on the best fitting are consistent with the upper bound ($g<6.6\times10^{-11}\  {\rm GeV}^{-1}$, i.e. $g_{11}<6.6$) set by the CAST experiment (\citealt{Anastassopoulos2017}). Comparing the fitting results of the two different models, we can find that the best-fitting $g_{11}$ which comes from the linearly-continuous-$\varphi$ model ($g_{11}$=0.7) is almost one order of magnitude smaller than that comes from the discrete-$\varphi$ model ($g_{11}$=5). This means that the coupling between photon and ALP in the linearly-continuous-$\varphi$ magnetic field structure is much weaker than in the discrete-$\varphi$ magnetic field structure, but the physical mechanism involved is still unclear.

\section{SUMMARY AND DISCUSSION}
ALPs are one promising kind of dark matter candidate particles that are predicted to couple with photons in the presence of magnetic fields. An ALP can oscillate into a photon and vise versa due to this coupling process. Such photon-ALP oscillations have been used to interpret observations of TeV gamma-ray photon from extragalactic sources, which are unexpected due to the electron-positron pair production process. In this paper, we obtain some new constraints on ALP properties based on photon-to-ALP conversion probability through TeV photons detected from a distant AGN PKS 2155-304.

First, we fit the broadband SED of PKS 2155-304 with a one-zone SSC model. Based on the fitting of its broadband SED, we can obtain the strength of the magnetic field $B=1.2\ {\rm G}$ around PKS 2155-304, and the survival probabilities of these Tev photons. With further reasonably assuming that the coherent domain size around the source is the distance from the central black hole to BLR (e.g., 1 pc); the strength of the intergalactic magnetic field (e.g., 1 nG), and the intergalactic coherent domain size (e.g., 1 Mpc), we can constrain the two key parameters for ALP, i.e., the particle mass $m_{a}$ and the photon-ALP coupling constant $g$ based on the survival probability of Tev photons. Two magnetic field configurations are considered based on the previous studies. One is the discrete-$\varphi$ model. In this model, the path of propagation is divided into lots of coherent domains, each has a uniform magnetic field and the same size $l$, the magnetic field has the orientation angle $\varphi$ changing discretely and randomly from one domain to the next. Another is the linearly-continuous-$\varphi$ model that the magnetic field orientation angle $\varphi$ varies continuously across neighboring domains.

For the discrete-$\varphi$ model, when $m_{1}$=0.1 and $g_{11}$=5 $(m_{1}\equiv m_{\rm a}/1\ {\rm neV}$, and $g_{11}\equiv g/10^{-11}\ {\rm GeV^{-1}})$, we can obtain the best fitting curve of $P_{\rm S,G}$. The reasonable range of the ALP mass $m_{1}$ is 0.08 $\thicksim$ 0.2 when $g_{11}$=5, and the only reasonable value of the photon-ALP coupling constant is $g_{11}$=5 when $m_{1}$=0.1. These results imply that, in the energy range $10^{25} -10^{27}\ {\rm Hz}$, for the discrete-$\varphi$ model, the Tev photon survival probabilities $P_{\rm S,G}$ are very sensitive to the ALP mass $m_{1}$, but are not sensitive to the photon-ALP coupling constant $g_{11}$.

For the linearly-continuous-$\varphi$ model, when $m_{1}$=0.1 and $g_{11}$=0.7, we can obtain the best fitting curve of $P_{\rm S,W}$. The reasonable range of the ALP mass $m_{1}$ is 0.05 $\thicksim$ 0.4 when $g_{11}$=0.7, and the reasonable range of the photon-ALP coupling constant $g_{11}$ is 0.5 $\thicksim$ 1 when $m_{1}$=0.1. These results imply that, in the energy range $10^{25} -10^{27}\ {\rm Hz}$, for the linearly-continuous-$\varphi$ model, the Tev photon survival probabilities $P_{\rm S,W}$ are very sensitive to the ALP mass $m_{1}$, and are also very sensitive to the photon-ALP coupling constant $g_{11}$.

It is notable that, for the two models, the best-fitting parameters of ALPs and the reasonable ranges of the parameters of ALPs which are based on the best fitting are consistent with the upper bound ($g<6.6\times10^{-11}\  {\rm GeV}^{-1}$, i.e. $g_{11}<6.6$) set by the CAST experiment. Compare the fitting results of the two different models, we can find that the best-fitting $g_{11}$ which comes from the linearly-continuous-$\varphi$ model ($g_{11}$=0.7) is almost one order of magnitude smaller than that comes from the discrete-$\varphi$ model ($g_{11}$=5). This means that the coupling between photon and ALP in the linearly-continuous-$\varphi$ magnetic field structure is much weaker than in the discrete-$\varphi$ magnetic field structure, but the physical mechanism involved is still unclear.

Although PKS 2155-304 is a well-studied TeV gamma-ray emitter, the Tev observations we can obtain are still limited. More publicly available Tev observations of PKS 2155-304 are necessary to obtain more precise constraints on ALP properties.

\begin{acknowledgements}
We thank Dr. Feng Huang for useful discussions and comments, which helped to improve the manuscript.
\end{acknowledgements}

\label{lastpage}


\begin{thebibliography}{99}
\bibitem[Anastassopoulos et al.(2017)]{Anastassopoulos2017} Anastassopoulos, V., Aune, S., Barth, K., et al.\ 2017, Nature Physics, 13, 584
\bibitem[Ade et al.(2015)]{Ade2015} Ade, P.~A.~R., Arnold, K., Atlas, M., et al.\ 2015, \prd, 92, 123509
\bibitem[Abramowski et al.(2013)]{Abramowski2013} Abramowski, A., Acero, F., Aharonian, F., et al.\ 2013, \prd, 88, 102003
\bibitem[Aharonian et al.(2005)]{Aharonian2005b} Aharonian, F., Akhperjanian, A.~G., Bazer-Bachi, A.~R., et al.\ 2005, \aap, 442, 895
\bibitem[Aharonian et al.(2005)]{Aharonian2005a} Aharonian, F., Akhperjanian, A.~G., Aye, K.-M., et al. 2005, \aap, 430, 865
\bibitem[Aharonian et al.(2007)]{Aharonian2007} Aharonian, F., Akhperjanian, A.~G., Bazer-Bachi, A.~R., et al.\ 2007, \apjl, 664, L71
\bibitem[Aharonian et al.(2009)]{Aharonian2009b} Aharonian, F., Akhperjanian, A.~G., Anton, G., et al.\ 2009, \aap, 502, 749
\bibitem[Aharonian et al.(2009)]{Aharonian2009a} Aharonian, F., Akhperjanian, A.~G., Anton, G., et al.\ 2009, \apjl, 696, L150
\bibitem[Ackermann et al.(2012)]{Ackermann2012} Ackermann, M., Ajello, M., Allafort, A., et al.\ 2012, Science, 338, 1190
\bibitem[Aharonian et al.(2006)]{Aharonian2006} Aharonian, F., Akhperjanian, A.~G., Bazer-Bachi, A.~R., et al.\ 2006, \nat, 440, 1018
\bibitem[Burrage et al.(2009)]{Burrage2009} Burrage, C., Davis, A.-C., \& Shaw, D.~J.\ 2009, Physical Review Letters, 102, 201101
\bibitem[Bloom \& Marscher(1996)]{Bloom1996} Bloom, S.~D., \& Marscher, A.~P.\ 1996, \apj, 461, 657
\bibitem[Balokovi{\'c} et al.(2016)]{Balokovi2016} Balokovi{\'c}, M., Paneque, D., Madejski, G., et al.\ 2016, \apj, 819, 156
\bibitem[Blumenthal \& Gould(1970)]{Blumenthal1970} Blumenthal, G.~R., \& Gould, R.~J.\ 1970, Reviews of Modern Physics, 42, 237
\bibitem[Cs{\'a}ki et al.(2002)]{Csaki2002} Cs{\'a}ki, C., Kaloper, N., \& Terning, J.\ 2002, Physical Review Letters, 88, 161302
\bibitem[Cs{\'a}ki et al.(2015)]{Csaki2015} Cs{\'a}ki, C., Kaloper, N., \& Terning, J.\ 2015, \jcap, 6, 041
\bibitem[Chen(2017)]{Chen2017} Chen, L.\ 2017, \apj, 842, 129
\bibitem[Dias et al.(2014)]{Dias2014} Dias, A.~G., Machado, A.~C.~B., Nishi, C.~C., Ringwald, A., \& Vaudrevange, P.\ 2014, Journal of High Energy Physics, 6, 37
\bibitem[Dermer \& Schlickeiser(1993)]{Dermer1993} Dermer, C.~D., \& Schlickeiser, R.\ 1993, \apj, 416, 458
\bibitem[Falomo et al.(1993)]{Falomo1993} Falomo, R., Pesce, J.~E., \& Treves, A.\ 1993, \apjl, 411, L63
\bibitem[Grossman et al.(2002)]{Grossman2002} Grossman, Y., Roy, S., \& Zupan, J.\ 2002, Physics Letters B, 543, 23
\bibitem[Grasso \& Rubinstein(2001)]{Grasso2001} Grasso, D., \& Rubinstein, H.~R.\ 2001, \physrep, 348, 163
\bibitem[Horns et al.(2012)]{Horns2012} Horns, D., Maccione, L., Meyer, M., et al.\ 2012, \prd, 86, 075024
\bibitem[H.E.S.S.~Collaboration et al.(2013)]{H.E.S.S.2013} H.E.S.S.~Collaboration, Abramowski, A., Acero, F., et al.\ 2013, \aap, 550, A4
\bibitem[H.E.S.S.~Collaboration et al.(2010)]{Abramowski2010} H.E.S.S.~Collaboration, Abramowski, A., Acero, F., et al.\ 2010, \aap, 520, A83
\bibitem[Jaeckel \& Ringwald(2010)]{Jaeckel2010} Jaeckel, J., \& Ringwald, A.\ 2010, Annual Review of Nuclear and Particle Science, 60, 405
\bibitem[Kataoka et al.(1999)]{Kataoka1999} Kataoka, J., Mattox, J.~R., Quinn, J., et al.\ 1999, \apj, 514, 138
\bibitem[Kuo \& Pantaleone(1989)]{Kuo1989} Kuo, T.~K., \& Pantaleone, J.\ 1989, Reviews of Modern Physics, 61, 937
\bibitem[Mirizzi et al.(2008)]{Mirizzi2008} Mirizzi, A., Raffelt1, G.~G., \& Serpico, P.~D.\ 2008, Axions, 741, 115
\bibitem[Mirizzi et al.(2005)]{Mirizzi2005} Mirizzi, A., Raffelt, G.~G., \& Serpico, P.~D.\ 2005, \prd, 72, 023501
\bibitem[Mirizzi \& Montanino(2009)]{Mirizzi2009} Mirizzi, A., \& Montanino, D.\ 2009, \jcap, 12, 004
\bibitem[Mazin \& Raue(2007)]{Mazin2007} Mazin, D., \& Raue, M.\ 2007, \aap, 471, 439
\bibitem[MAGIC Collaboration et al.(2008)]{MAGIC2008} MAGIC Collaboration, Albert, J., Aliu, E., et al.\ 2008, Science, 320, 1752
\bibitem[Maraschi et al.(1992)]{Maraschi1992} Maraschi, L., Ghisellini, G., \& Celotti, A.\ 1992, \apjl, 397, L5
\bibitem[{\"O}stman \& M{\"o}rtsell(2005)]{Ostman2005} {\"O}stman, L., \& M{\"o}rtsell, E.\ 2005, \jcap, 2, 005
\bibitem[Peccei \& Quinn(1977)]{Peccei1977} Peccei, R.~D., \& Quinn, H.~R.\ 1977, Physical Review Letters, 38, 1440
\bibitem[Rybicki \& Lightman(1979)]{Rybicki1979} Rybicki, G.~B., \& Lightman, A.~P.\ 1979, New York, Wiley-Interscience, 1979.~393 p.
\bibitem[Simet et al.(2008)]{Simet2008} Simet, M., Hooper, D., \& Serpico, P.~D.\ 2008, \prd, 77, 063001
\bibitem[Smith et al.(1995)]{Smith1995} Smith, E.~P., O'Dea, C.~P., \& Baum, S.~A.\ 1995, \apj, 441, 113
\bibitem[Sikora et al.(1994)]{Sikora1994} Sikora, M., Begelman, M.~C., \& Rees, M.~J.\ 1994, \apj, 421, 153
\bibitem[S{\'a}nchez-Conde et al.(2009)]{Conde2009} S{\'a}nchez-Conde, M.~A., Paneque, D., Bloom, E., Prada, F., \& Dom{\'{\i}}nguez, A.\ 2009, \prd, 79, 123511
\bibitem[Wang \& Lai(2016)]{Wang2016} Wang, C., \& Lai, D.\ 2016, \jcap, 6, 006
\end{thebibliography}
\end{document}